\documentclass[
  reprint,
 superscriptaddress,
bibnotes, 
 aps,
pra,
]{revtex4-2}
\usepackage{orcidlink}
\usepackage{graphicx}
\usepackage{physics}
\usepackage{xspace}
\newcommand{\ps}{phase space\xspace}
\newcommand{\state}{{\hat {\varrho}}}
\usepackage{slashbox}
\usepackage{makecell}
\usepackage{multirow}
\usepackage{diagbox}

\usepackage{color}
\usepackage{soul}
\sethlcolor{yellow}
\soulregister\cite7
\soulregister\ref7
\soulregister\pageref7
\soulregister\flalign7
\soulregister\xspace7
\soulregister\ps7

\newcommand{\VEC}[1]{{\mbox{\boldmath${#1}$}}}

\usepackage[normalem]{ulem}

\def\W0{W^{0}}
\def\Wt{W^{\tau}}

\def\rV{{\VEC{r}}}
\def\rVf{{\VEC{r}^\prime}}

\def\xf{{{x}^\prime}}
\def\pf{{{p}^\prime}}

\usepackage{hyperref}

\begin{document}

\title{Monitoring Beam Splitter Entanglement using Quantumness}

\author{Hua-Li Chen} 
\affiliation{Department of Physics, National Tsing Hua University, Hsinchu 30013, Taiwan}

\author{Hsien-Yi Hsieh\orcidlink{0000-0001-5227-8248}}
\affiliation{Institute of Photonics Technologies, National Tsing Hua University, Hsinchu 30013, Taiwan}

\author{Chien-Ming Wu} 
\affiliation{Institute of Photonics Technologies, National Tsing Hua University, Hsinchu 30013, Taiwan}

\author{Ole Steuernagel\orcidlink{0000-0001-6089-7022}}
\affiliation{Institute of Photonics Technologies, National Tsing Hua University, Hsinchu 30013, Taiwan}

\author{Ray-Kuang Lee\orcidlink{0000-0002-7171-7274}}
\email{rklee@ee.nthu.edu.tw}
\affiliation{Department of Physics, National Tsing Hua University, Hsinchu 30013, Taiwan}
\affiliation{Institute of Photonics Technologies, National Tsing Hua University, Hsinchu 30013, Taiwan}
\affiliation{Center for Theory and Computation, National Tsing Hua University, Hsinchu 30013, Taiwan}
\affiliation{Center for Quantum Science and Technology, Hsinchu 30013, Taiwan}
 
\date{\today}
\begin{abstract}
  We report on an experiment in which two independent squeezed vacuum states get entangled by mixing
  them with a balanced beam splitter. We follow standard practice and use an inseparability
  criterion to quantify their entanglement.  However, this only allows us to witness the
  entanglement, but not to determine the deleterious effects of experimental imperfections due to
  the beam splitter mixing and the associated mode-mismatch and detection imperfections. We
  therefore introduce an alternative framework suitable for continuous variable systems using the
  states' {\it quantumness, $\Xi$}. We show that, under ideal circumstances, $\Xi$ is a conserved
  quantity under beam mixing. This allows us to benchmark the experiment's performance by comparing
  the states' quantumness~$\Xi$ after the beam splitter mixing with $\Xi$ before. Such a comparison is not
  possible with entanglement witnesses, as the input states are unentangled. This highlights the main
  strength of our approach: its ability to \emph{generally quantify} the quantumness of multi-mode
  continuous variable states and use this to probe different stages in an experiment.
\end{abstract}

\maketitle

Typically, the non-classical features of quantum states are fragile with respect to experimental
imperfections. Here we introduce a state's quantumness, $\Xi$~\cite{Ole_23_Quantumness}, as a tool
to quantify experimental performance by monitoring the non-classical character of a state in
continuous-variable multi-mode~\cite{Ole_25_TowardsQuantumness} (bosonic) systems.

\begin{figure}[b] \centering
\includegraphics[width=8.4cm,height=3.2cm]{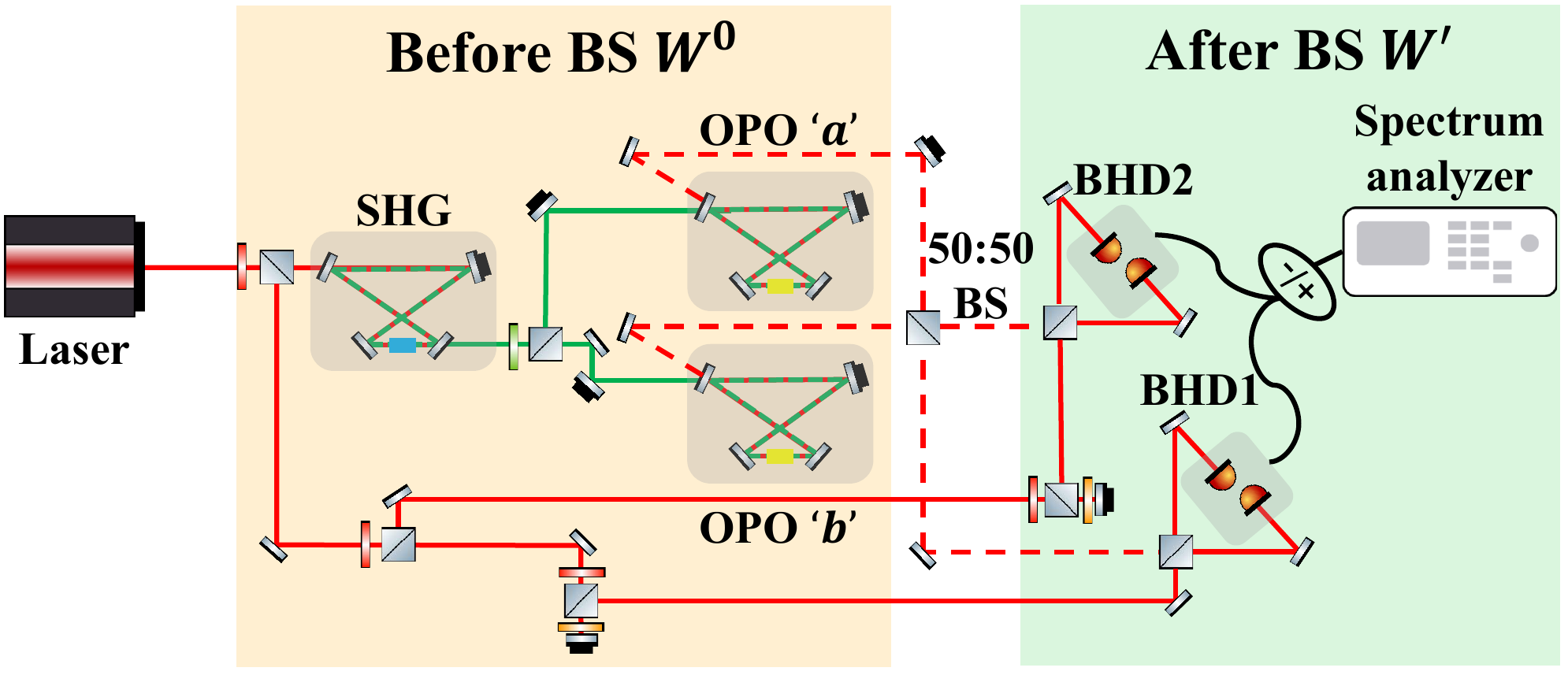}
\caption{In \mbox{our experiment two degenerate Optical Parametric} Oscillators (OPOs),
  each generate single-mode squeezed va\-cu\-um states with orthogonal squeezing
  angles, serving  as `input states', $W^0$, for
   a balanced (50\%:50\%) beam splitter (BS), yielding  entangled output states, $W^\prime$, see Fig.~\ref{Fig:BS}.
   These are characterized by balanced homodyne detection (BHD) fed either separately
   or jointly into a spectrum analyzer.
   \label{Fig:HuaLi_Experiment}}
\end{figure}

In our experiment, see Fig.~\ref{Fig:HuaLi_Experiment}, we employ a quantum-optical setup, mixing
independent squeezed-vacuum states in two modes at a balanced beam splitter. This, inevitably,
creates losses and reduces the quantum correlations across the modes.  

Therefore, instead of
following the customary route of using inseparability criteria which, for unentangled input states,
are `blind' to the states' quantum characteristics before the beam splitter, we characterize the
performance of our experiment by determining the states' quantumness~\cite{Ole_23_Quantumness}
\emph{before and after} the beam splitter.

The ability of quantumness~$\Xi$~\cite{Ole_23_Quantumness} to provide for such comparisons is the
central motivation for its use in this work.

Additionally, for the squeezed states we use here, quantumness~$\Xi$ faithfully discriminates
non-classical from classical states, because they are gaussian
states~\cite{Ole_23_Quantumness,Ole_25_TowardsQuantumness}.

Moreover, for any ideal mode-matched
lossless beam splitter all quantum-states' quantumness~$\Xi$ is always
conserved~\cite{Ole_26_BeamSplitter4D} and therefore all reductions in $\Xi$ represent losses of the
states' non-classical features directly attributable to experimental imperfections.

\begin{figure}[b] \centering
\includegraphics[width=2.9cm, height=3.2cm]{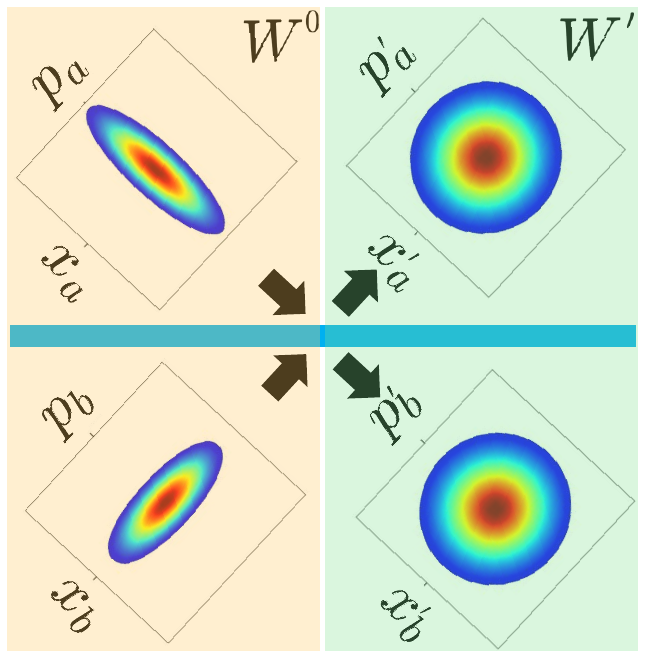}
\includegraphics[width=5.6cm,height=3.2cm]{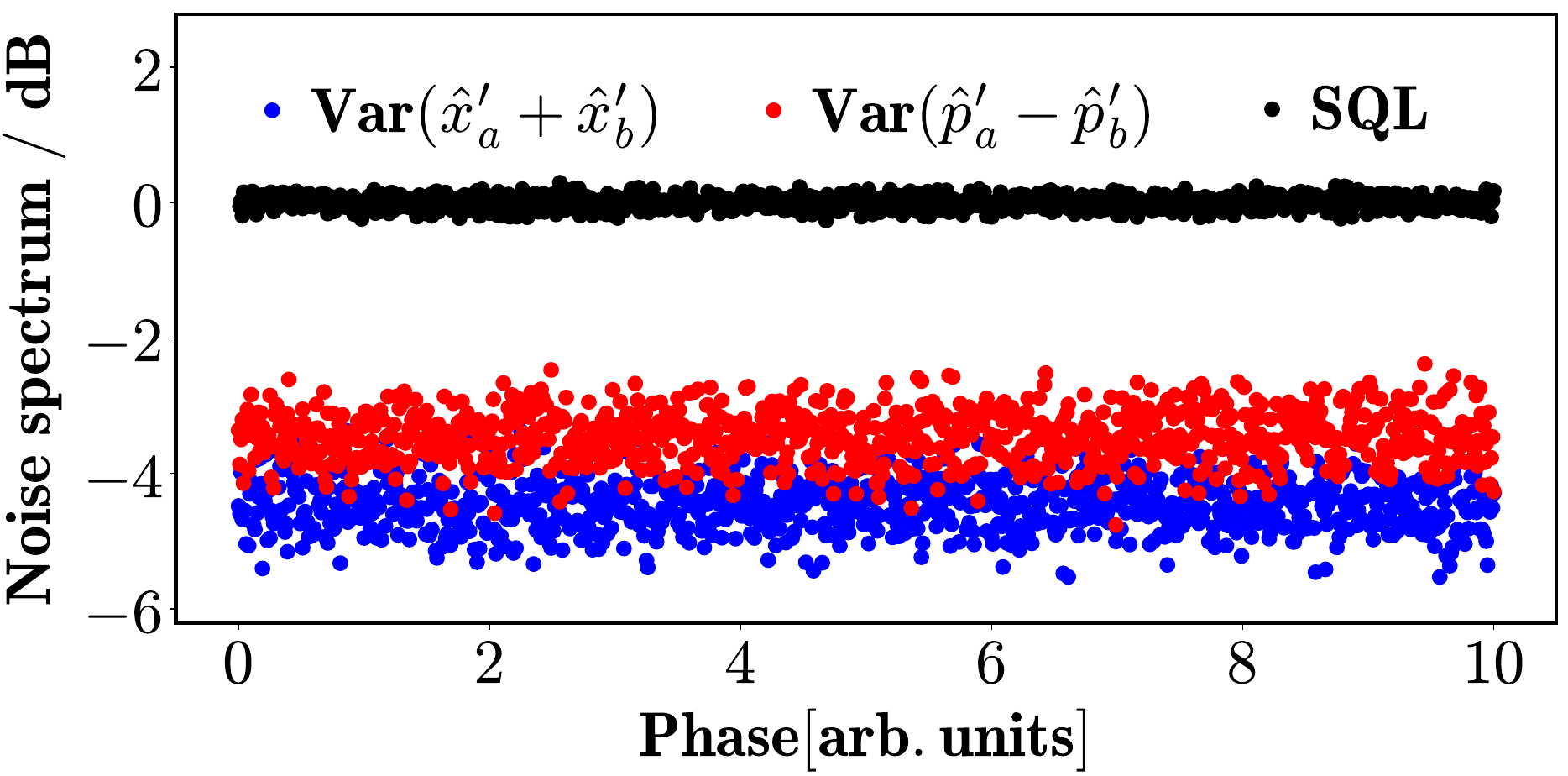}
\caption{The squeezed quadratures of the input states,~$W^0$, have variances below the vacuum's
  standard quantum limit (SQL), but they are above the SQL in the reduced
  output states~$W^\prime_{I}$~(\ref{eq:ClassicalStateafterBS}). Yet, the below-SQL fluctuations
  are not lost,  the beam splitter has  mapped them into reduced cross-mode
  variances Var$(\hat x_{+}^\prime)$ and Var$(\hat p_{-}^\prime)$, see  Eqs.~(\ref{eq:stateAfterBS2_MinusPlus}) and~(\ref{eq:_Duan_C}).
  \label{Fig:BS}}
\end{figure}

Here we study the EM-field fluctuations of quantum-optical squeezed
states~\cite{Slusher_Hollberg__PRL85__SqueeObserv,Levenson_Reid_Walls__PRL86_Quadrature,
  Ou_Mandel__PRL88_BellViol,Ou_Zou_Wang_Mandel__PRL90__BellObserv,Scully_Zubairy__Book01,Lvovsky_16squeezed}. We
use two independent squeezed vacuum states in two matched modes, see
Figs.~\ref{Fig:HuaLi_Experiment} and~\ref{Fig:BS}, each mode displaying non-classical fluctuations
--below those of the vacuum state~\cite{Wuensche_JOBQSO04}.

In quantum optics beam splitters can be used to entang\-le such
modes~\cite{Titulaer_Glauber__PR66,HongOuMandel87}, in our case this means that the non-classical
fluctuations are not present in either individual mode after mixing by the beam splitter, see
Fig.~\ref{Fig:BS} and Eq.~(\ref{eq:ClassicalStateafterBS}) below (also, Fig.~3 of
Ref.~\cite{Ekert_Knight__AMJP89}). Instead, they are spread out across both
modes~\cite{Ekert_Knight__AMJP89,Ekert_Knight__AJP95__EntangSchmidt,Lvovsky_16squeezed,Braunstein__PRA05__SqueezResource}. This
way the non-classical nature of below vacuum-level fluctuations~\cite{Wuensche_JOBQSO04} (below the
`Standard Quantum Limit (SQL)~\cite{Lvovsky_16squeezed}) of our two squeezed-vacuum input states
seem to have been lost post-beam
splitter~\cite{Ekert_Knight__AMJP89,Ekert_Knight__AJP95__EntangSchmidt}. But they can still be
retrieved using joint measurements of quadrature operators across both modes, see Fig.~\ref{Fig:BS}
and Ref.~\cite{Cerf_Leuchs_Polzik_2007}.

In this way we end up with distributions that mimic the original scheme by Einstein, Podolsky and
Rosen (EPR)~\cite{Einstein_Podolski_PR35} which have found various applications~\cite{EPR-comb,
  EPR-W, EPR-cloud}.  They exhibit strong non-classical correlations in their complementary
observables~\cite{Ou_Pereira_Kimble__PRL92,CV-entangle} (however, see Bell's cautioning
remarks~\cite{Ou_Pereira_Kimble__PRL92} on classical descriptions of the standard
EPR-scheme~\cite[Chapter~21]{Bell__Book04__SpeakUnspeak}).

Todays' customary approach uses inseparability criteria such
as~\cite{Duan_Cirac_Zoller__PRL00_inseparability,Laurat_2005, CV-QIP} to quantify the success in
generating non-classical cross-mode correlations (such as reported in Fig.~\ref{Fig:BS}) and the
associated entanglement.

Here, we apply Duan's  inseparability certification criterion~\cite{Duan_Cirac_Zoller__PRL00_inseparability},
which is based on constraints for the sum of variances of the joint quadratures. For our case (see
Fig.~\ref{Fig:BS}) it takes the form~\cite{Eberle2013, Yu2014}
\begin{eqnarray}\label{eq:_Duan_C}
  D_C \equiv \text{Var}(\hat{x}^\prime_a + \hat{x}^\prime_b)
  + \text{Var}(\hat{p}^\prime_a - \hat{p}^\prime_b) < 2 \; .
\end{eqnarray}

\begin{figure}[b] \centering
 \includegraphics[width=7cm]{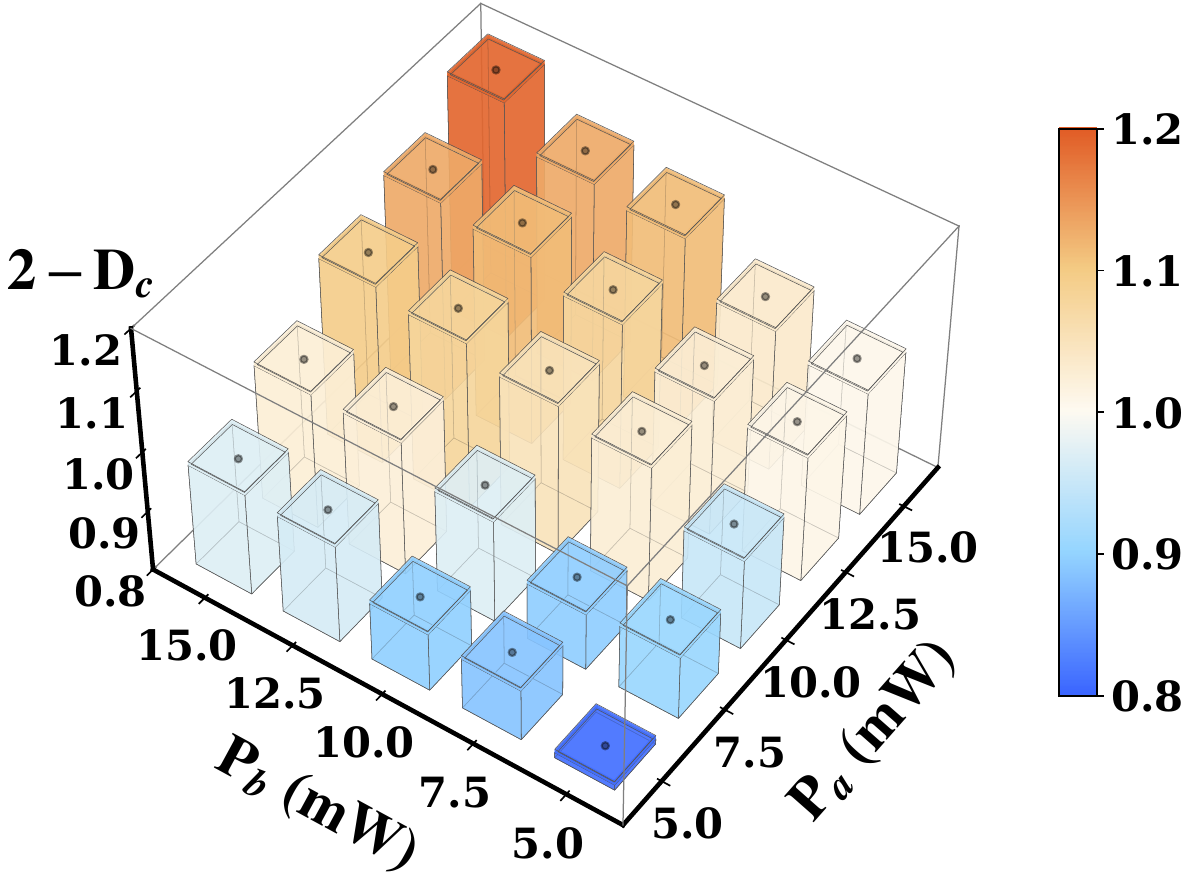}
 \caption{Positive values of $2$-$D_C$, for Duan's criterion $D_C$ of Eq.~(\ref{eq:_Duan_C}),
   certify cross mode entanglement for all OPO-power settings, see Fig.~\ref{Fig:HuaLi_Experiment}, of our experiment.
   Table~\ref{tab:duan_values} contains the experimentally determined values displayed here. 
   \label{Fig:Duan}}
\end{figure}

We performed joint quadrature measurements after the beam splitter and used Duan's criterion to certify and quantify
mode entanglement for all settings of our experiment, see Fig.~\ref{Fig:Duan}.

In experiments we typically start out with separable input states. Consequently, inseparability
criteria do not allow us to monitor experimental performance very well: inseparability criteria are
`blind' to the non-classical features of the separable input states. They are designed to only `see'
the post-beam splitter entanglement. That is why we now consider quantumness~$\Xi$ instead of separability.

Quantumness~$\Xi$ is a \ps-based measure and always conserved under perfect beam splitter
operations~\cite{Ole_26_BeamSplitter4D}. In our experiment all reductions in $\Xi$ represent losses
associated with the beam splitter operation and associated experimental imperfections.  Gaussian
states of continuous variable systems, in atom~\cite{Kurtsiefer_NAT97,Scully_Zubairy__Book01} as
well as quantum optics~\cite{Scully_Zubairy__Book01} are conveniently
reconstructed~\cite{Hsieh__PRL22}, studied and depicted as objects in \ps, see Fig.~\ref{Fig:BS}.

For our purposes it is convenient to use Wigner's phase space distribution,~$W$, as a starting
point, It is generated from the two-mode density operator ${\varrho(x,z;x^\prime,z^\prime)}$
$ = \langle x,z | \hat \varrho |x^\prime,z^\prime \rangle$ by (two) modewise Fourier-transformations
\begin{flalign}\label{eq:_WignerDistr}
  W(x_a,p_a,&x_b,p_b) =  \frac{1}{\pi^2} \iint_{-\infty}^{\infty} dy_a dy_b
  \; {\rm e}^{{2 {\rm i}} \,p_a y_a} {\rm e}^{{2 {\rm i}} \, p_b y_b } \notag \\
 & \times \; \varrho(x_a-y_a,x_b-y_b;x_a+y_a,x_b+y_b),
\end{flalign}
with respect to the distances~$y$ spanned by its off-diagonal
coherences~\mbox{\cite{Wigner_PR32,Hillery_PR84}}.  By construction, $W$ is a distribution of
position~$x$ and momentum~$p$ in \ps, $W$ is normalized and nonlocal (through~$y$). Whereas~$\varrho$
tends to be complex-valued, $W$ is always real-valued but, generi\-cally, $W$ has nega\-tive values
in some regions of \ps~\cite{Wigner_PR32,Hillery_PR84}. Since~$W$ and~$ \varrho$ are Fourier
transforms~(\ref{eq:_WignerDistr}) of each other, they are unitarily equivalent to each other.

Associated with $W$ is Husimi's $Q$-distribution, a smeared version
of~$W$~\cite{Cahill_PR69b,Schleich_01}, generated by convolution with the vacuum state:
\begin{flalign}
  Q (x,p) = \iint_{-\infty}^{\infty} \! dp \, dx \; \frac{1}{\pi} \; {{\rm e}^{-x^2 -p^2}} W(x,p) \;
  ,
  \label{eq:Husimi_Q} 
\end{flalign}
and convolution with $\tfrac{1}{\pi^2}  {{\rm e}^{-x_a^2 -p_a^2-x_b^2 -p_b^2}} $ for the two modes
case $ Q(x_a,p_a,x_b,p_b)$.

Bohmann and Agudelo devised~\cite{Bohmann_PRL20} a certification functional,~$\xi[W]$, for the presence
of nonclassical behaviour of a state $\state$ based  on $W$ and~$Q$
\begin{flalign}
  & \xi[W](x,p) =  W(x,p) - 4 \pi \; Q^2(x,p) \; .
  \label{eq:xi_Certification} 
\end{flalign}
$\xi$ is the most sensitive certification functional we could find~\cite{Ole_23_Quantumness,Ole_25_TowardsQuantumness}: $\xi[W](x,p) < 0 $ certifies
that a state~$W$ is nonclassical; if $W$ is classical, then $\xi \geq 0$ throughout
\ps~\cite{Bohmann_PRL20}.  There are, however, weakly nonclassical states for which $\xi \geq 0$:
in other words, $\xi$ does
not always faithfully discriminate between classical and quantum states, but for gaussian states,
such as the squeezed states considered here, $\xi$ is always fully faithfully
discriminating~\cite{Ole_25_TowardsQuantumness}.

This led us to construct, $\Xi$, a measure of {\it quantumness} of a state~\cite{Ole_23_Quantumness}
\begin{flalign}
  \Xi[W] = \iint_{-\infty}^{\infty} dx \; dp \;\; \big.\VEC{\Delta} \xi[W] (x,p)\Big|_{\xi < 0} \; ,
  \label{eq:Xi_Measure} 
\end{flalign}
where,
$\VEC{\Delta} \xi = \frac{\partial^2 \xi}{\partial x^2} + \frac{\partial^2 \xi}{\partial p^2}$ is
the \ps Laplacian and our quantum\-ness measure only includes areas in \ps (`basins' where
$\xi(x,p) <0$) where a state displays non\-classicality.  In our experiment, as the input states are separable, all states can be
expressed or re-expressed in product form~\cite{Ole_26_BeamSplitter4D}. We can therefore, despite
the fact that we investigate a two-mode system, restrict ourselves to this single-mode
formulation~(\ref{eq:Xi_Measure}) of our quantumness measure.

In our experiment, we create two impure squeezed states $W_a$ and $W_b$ using squeezers `$a$' and
`$b$', see Fig.~\ref{Fig:HuaLi_Experiment}. Therefore, the initial state $\W0$ is described by their product, i.e.,
\begin{flalign}
\!\!\!  W^{0}\! = \!
  W_a W_b = \!\! \frac{\exp \left[-\frac{p_a^2}{V_{p_a}}-\frac{x_a^2}{V_{x_a}}\right]}{\pi
    \sqrt{V_{p_a} V_{x_a}}} \frac{\exp \left[
      -\frac{p_b^2}{V_{x_b}}-\frac{x_b^2}{V_{p_b}}\right]}{\pi \sqrt{V_{p_b} V_{x_b}}},
 \! \label{eq:W0_productState}
\end{flalign}
where the respective spreads~$V$ (which are twice the variances) have to obey Heisenberg's
uncertainty principle which in this language takes the form $ V_{x_\#} V_{p_\#} \geq 1$ for either
mode `$\#$'$ = a$ or $b$. Eq.~(\ref{eq:W0_productState}) is the generalization of the well-known
pure state result~\cite{Ekert_Knight__AMJP89} to mixed states.

Note that in state $W_b$ we label the spread~$V$ in coordinate $p_b$ by $V_{x_b}$ and that in $x_b$
by $V_{p_b}$ in order to emphasize that we use a `crossed' configuration (orthogonal in their
squeezing angles): the squeezed state in mode `$b$' is rotated by $90^\circ$ with respect to the
state in mode `$a$'. In this crossed configuration we assign the coordinates $p_a$ and $x_b$ of the
respective modes `$a$' and `$b$' as the `narrowly squeezed' directions: namely,
$\exp[ -\frac{p_a^2}{V_{p_a}} ] = \exp[ -{V_{a} p_a^2} ]$, with spread $V_{p_a} \leq 1$ or inverse
spread $V_{a} \geq 1$, and, analogously, $\exp[ -\frac{x_b^2}{V_{p_b}} ] = \exp[ -{V_{b} x_b^2} ]$
with $V_{b} \geq 1$, see Fig.~\ref{Fig:BS}.

If mode `$a$' were in a pure squeezed state then {$ V_{x_a} V_{p_a} = V_{a} \!\times\! 1/V_{a} = 1$}
would saturate Heisenberg's uncertainty threshold, but in our case, for mode `$a$', the
anti-squeezed coordinate, $x_a$, carries the increased spread $ \mu_a V_{a} $ with an impurity
factor, $\mu_a > 1\!\!: V_{x_a} V_{p_a} \mapsto \mu_a V_{a} \!\times\!  1/V_{a} = \mu_a > 1$. Analogous
considerations apply to mode `$b$' yielding a spread of $\mu_b V_b$, for $p_b$.

\begin{figure}[h!] \centering
  \includegraphics[width=7.0cm]{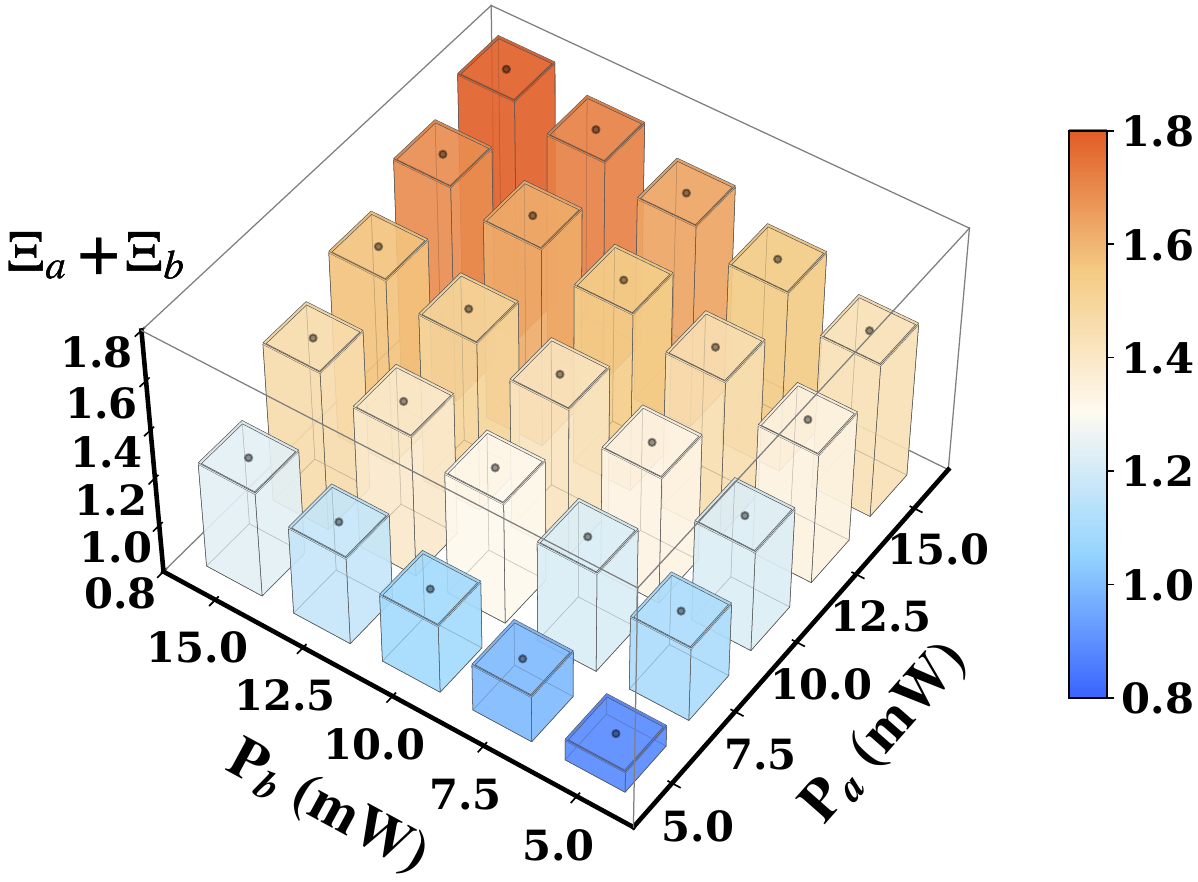}
  \caption{Quantumness, $\Xi_a + \Xi_b $, of modes before the beam splitter, rendered on OPO pump power grid, see Fig.~\ref{Fig:HuaLi_Experiment}.
    \label{Fig:XiGridPlotBefore}}
\end{figure}

Therefore, the initial state~(\ref{eq:W0_productState}) can be rewritten as the impure product state
\begin{flalign}
  W^{0} = \frac{\exp \left[- V_{a} p_a^2 -\frac{x_a^2}{\mu_a V_{a}}\right]}{\pi  \sqrt{\mu_a}} \frac{\exp \left[
      -V_{b} {x_b^2} - \frac{p_b^2}{\mu_b V_{b}}\right]}{\pi \sqrt{\mu_b}}
  \label{eq:W0_productState_V_mu} ,
\end{flalign}
and, owing to its product form, we can trace out one or the other state implying that the
quantumness of this state is $\Xi[W_a] + \Xi[W_b] = \Xi_a + \Xi_b$, see Figs.~\ref{Fig:XiGridPlotBefore}
and~\ref{Fig:input}.

If this state is fed into a balanced, perfect beam splitter, the state undergoes the coordinate
transformations~\cite{Leonhardt_Paul__PRA93,Ole_26_BeamSplitter4D}
\begin{flalign}
  \label{eq:_R_ps_explicit_5050_copied}
  \rVf(\rV, \tau=\frac{1}{2}) = \left(\!
    \begin{array}{c}
  {\xf_a} \\   
  {\pf_a} \\
  {\xf_b} \\
  {\pf_b} \\
    \end{array}\!\right) = \frac{1}{\sqrt{2}} \left(\!
    \begin{array}{c} {x_a} + {x_b}
      \\ {p_a} + {p_b} \\ {x_b}  - {x_a}  \\ {p_b}  - {p_a} 
    \end{array}\!\right) \; ,
\end{flalign}
with reflectance and transmittance  $r=\tau=\frac{1}{2}$.
The output state after the beam splitter becomes an entangled state of the form
\begin{eqnarray}
       W^\prime(\rVf) \! & = & \! \frac{1}{\pi^2 \sqrt{\mu _a \mu _b}} \exp \left[ - \frac{
          \left(\pf_a-\pf_b\right){}^2 }{2 \mu _b V_b} - \frac{ \left(\xf_a+\xf_b\right){}^2}{2 \mu _a V_a } \right] \nonumber\\
          &  \times & \!\! \exp \left[- \frac{ V_a \left(\pf_a+\pf_b\right){}^2+V_b
          \left(\xf_a-\xf_b\right){}^2 }{2 } 
      \right] .
      \label{eq:stateAfterBS2}
\end{eqnarray}

Tracing one mode [either `($\xf_a,\pf_a$)' or `($\xf_b,\pf_b$)'] out of $W^\prime$ yields reduced,
single-mode states, $W^\prime_I$, in terms of the remaining mode `($\xf,\pf$)', with
\begin{eqnarray}
  W^\prime_I (\xf,\pf) = \frac{{2 \exp \left[-\frac{2 \pf^2 V_a}{V_a \mu _b V_b+1}-\frac{2 \xf^2 V_b}{\mu _a
        V_a V_b+1}\right]}}{
      \left({\pi \sqrt{\mu _a V_a \mu _b V_b+\frac{1}{V_a V_b}+\mu _a+\mu
          _b}}\right)};
          \label{eq:ClassicalStateafterBS}
                   \end{eqnarray}
for symmetry reasons, the $W^\prime_I$ are of identical functional form for both remaining modes
[`($x^\prime_b,p^\prime_b$)' or `($x^\prime_a,p^\prime_a$)'].

With the values of $\mu,V \geq 1$, the spreads the $W^\prime_I$ have in both coordinates,
$(\xf,\pf)$, are greater than unity, namely, larger than vacuum fluctuations, see
Fig.~\ref{Fig:BS}. These states therefore by themselves have the form of thermal
states~\cite{Lvovsky_16squeezed}, and are classical~\cite{Wuensche_JOBQSO04} with zero
quantumness:~$\Xi[W^\prime_I]=0$~\cite{Bohmann_PRL20,Ole_23_Quantumness,Ole_25_TowardsQuantumness}.


However, combining the post-beam splitter coordinates into 
$\xf_{\pm} = (\xf_a \pm \xf_b)/\sqrt{2}$ and $\pf_{\pm} = (\pf_a \pm \pf_b)/\sqrt{2}$, we
effectively reverse the effects of the beam splitter~\cite{Ole_26_BeamSplitter4D} encoded by
Eq.~(\ref{eq:_R_ps_explicit_5050_copied}) and arrive at the expression
    \begin{eqnarray}
    \!\!\!\!\!  &\Wt(&\!\!\! \xf_-,\pf_-,\xf_+,\pf_+) = \Wt_{-}(\xf_-,\pf_-) \times \Wt_{+}(\xf_+,\pf_+) \;\; \nonumber\\
       & = & \!\!\! \frac{\exp \left[-V_a
          \pf_{-}^2-\frac{\xf_{-}^2}{\mu _a V_a} \right] \exp \left[-V_b
          \xf_{+}^2-\frac{\pf_{+}^2}{\mu _b V_b}\right]}{\pi ^2 \sqrt{\mu _a} \sqrt{\mu _b}}
\label{eq:stateAfterBS2_MinusPlus} .
\end{eqnarray}
Eq.~(\ref{eq:stateAfterBS2_MinusPlus}) formally unentangles the post-beam splitter state $W^\prime$~(\ref{eq:stateAfterBS2}),
allowing us, under ideal circumstances, to perfectly retrieve the input product-form of the
state~(\ref{eq:W0_productState_V_mu}), with coordinates~$x_a$ and $x_b$, in terms of post-beam
splitter measurements based on the coordinates $\xf_{\pm}$ and $\pf_{\pm}$, see
Fig.~\ref{Fig:HuaLi_Experiment}. This is a special case of the general rule that perfect beam
splitters induce stiff rotations in \ps~\cite{Leonhardt_Paul__PRA93,Ole_26_BeamSplitter4D}, which
implies conservation of $\Xi$ when states pass through ideal beam splitters: 
$ \Xi^0_{a,b} = \Xi^\prime_{ab}$.

Here, it also implies that $ \Xi[\Wt_{+}\Wt_{-}] = \Xi^\prime_{+} + \Xi^\prime_{-}$, with
$\Xi_a = \Xi^\prime_{-}$ and $\Xi_b = \Xi^\prime_+$.

Since $\Xi$ is conserved in the case of ideal beam splitters, any experimentally observed reductions
$\Xi_a > \Xi^\prime_{-}$ and $\Xi_b > \Xi^\prime_+$ allow us to monitor experimental losses incurred
due to, e.g., phase fluctuations, imperfect mode-matching and detection imperfections.

\begin{figure}[t] \centering
  \includegraphics[width=7.3cm]{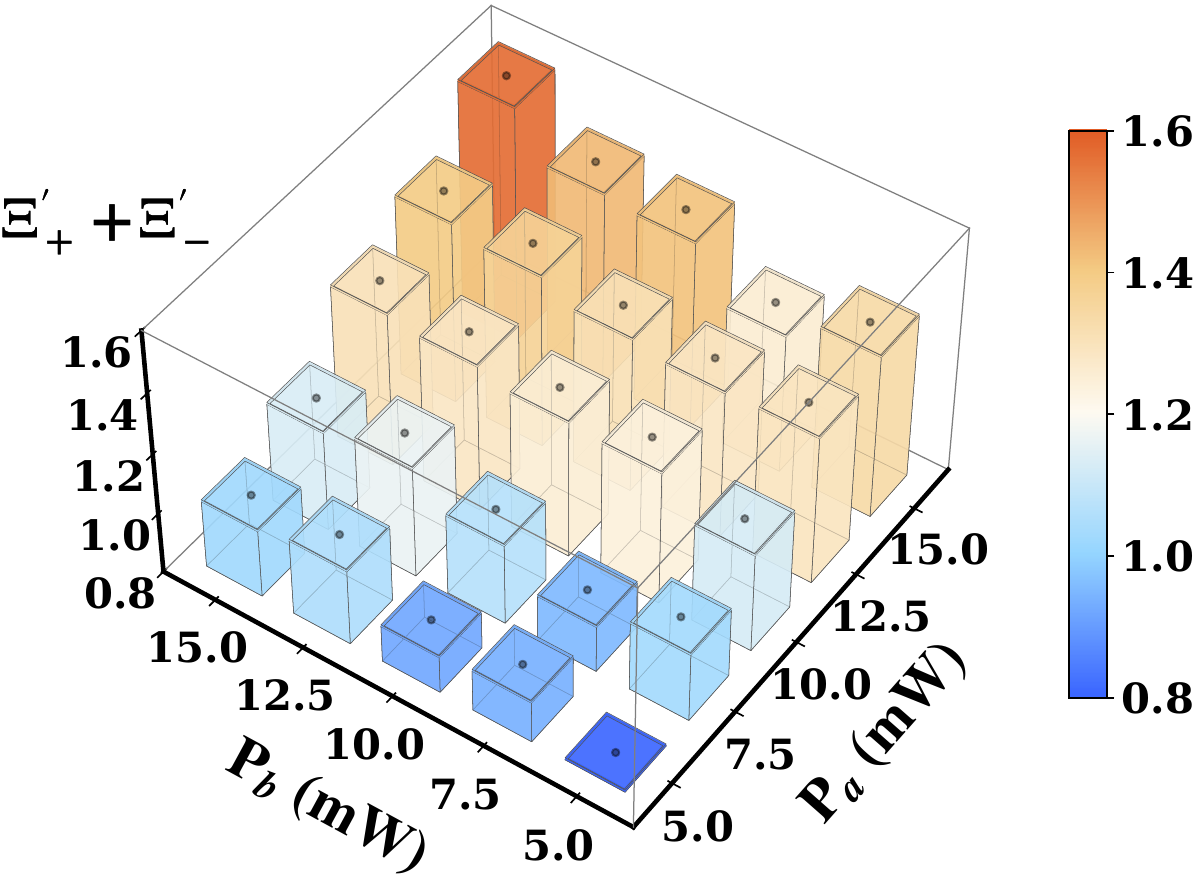}\\
  \includegraphics[width=7.3cm]{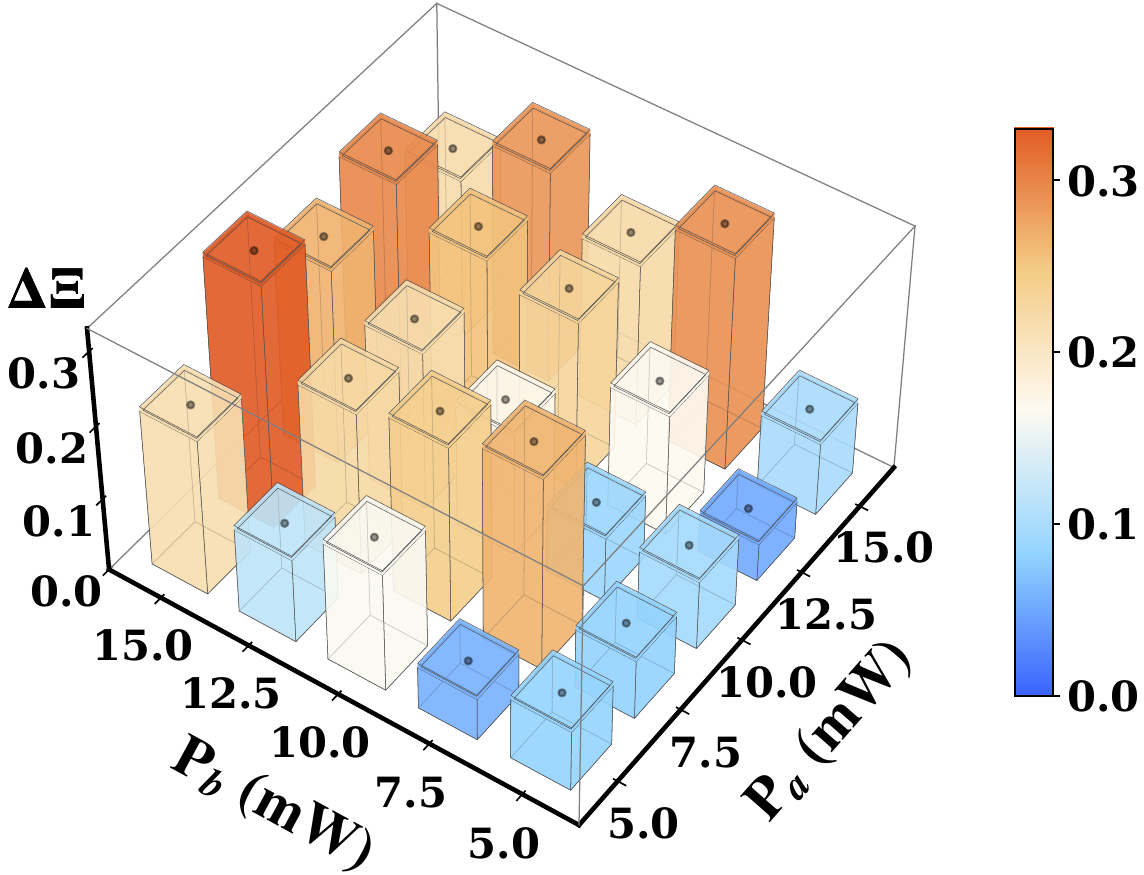}
  \caption{Quantumness, $\Xi^\prime_{+}+\Xi^\prime_{-}$, after
    the beam splitter, and their differences,
    $\Delta \Xi = (\Xi_a + \Xi_b) - (\Xi^\prime_{+}+\Xi^\prime_{-})$, rendered on OPO pump power
    grids, see Fig.~\ref{Fig:HuaLi_Experiment}. This rendering shows us that the
    experimental imperfections are random rather than due to systematic errors. Related experimental data are given in
    the Supplementary Information in tables~\ref{tab:opo_data}-~\ref{tab:detailed_joint_data}. 
    \label{Fig:XiGridPlots}}
\end{figure}

Inseparability criteria such as Duan's criterion do not allow us to compare experimental performance
in terms of `before' and `after' the mixing at the beam splitter; quantumness does. We
determined $\Xi^\prime_{+}+\Xi^\prime_{-}$, after the beam splitter, and the differences between
\emph{before} and \emph{after}, $\Delta \Xi = (\Xi_a + \Xi_b) - (\Xi^\prime_{+}+\Xi^\prime_{-})$,
see Fig.~\ref{Fig:XiGridPlots}.

We notice that the plot of $\Delta \Xi $ in Fig.~\ref{Fig:XiGridPlots} only shows positive
values. This behaviour conforms with our theoretical predictions since experimental imperfections
can only have degraded the states since no attempts have been made to use amplification techniques
to overcome losses~\cite{Bloomer__NJP11__SqueezRestore,OPA}, or distillation of entanglement~\cite{distill,
  distill-exp}. Therefore, the quantumness drops when passing the states through the beam splitter:
$\Xi_a + \Xi_b > \Xi^\prime_{+}+\Xi^\prime_{-}$.

Our rendering on the OPO power-grid, in Figs.~\ref{Fig:Duan}-~\ref{Fig:XiGridPlots}, additionally
allows us to see that the experimental imperfections are quite random rather than due to systematic
errors, allowing us to qualitatively characterize experimental performance as well as quantify and
display experimental performance in this new and versatile way.

Duan's entanglement criterion $2-D_C$ in Fig.~\ref{Fig:Duan}, behaves qualitatively similarly to the
quantumness, $\Xi_+ + \Xi_- $, of the modes after the beam splitter displayed in
Fig.~\ref{Fig:XiGridPlots}. And yet, Fig.~\ref{Fig:Duan} reveals that Duan's criterion is quite
considerably less sensitive than quantumness: Whereas, owing to experimental imperfections, the
quantumness of the states exiting from the beam splitter does not always monotonously rise with
increasing pump power, see Fig.~\ref{Fig:XiGridPlots}, Duan's criterion is too insensitive to pick
this up.  Visually, Fig.~\ref{Fig:Duan} behaves more like Fig.~\ref{Fig:XiGridPlotBefore} depicting
quantumness \emph{before} the beam splitter although Duan's criterion can only be applied to the
states \emph{after} the beam splitter. Just by itself, when applied to the beam splitter output
states, quantumness quantifies mode entanglement better than Duan's inseparability criterion, even
without using the contrast with the quantumness of the input states, as displayed in
Fig.~\ref{Fig:XiGridPlots}.

\emph{To conclude}, we experimentally demonstrated that experimental performance in a
quantum-optical entanglement experiment can be monitored and quantified using the quantum states'
quantumness.  This approach has great versatility, as we demonstrated when contrasting it with the
use of commonly applied inseparability criteria.  It is conceptually simple, generally applicable in
single and multi-mode settings~\cite{Ole_25_TowardsQuantumness}. It moreover has great
sensitivity~\cite{Ole_25_TowardsQuantumness} and is agnostic regarding details: it is applicable
irrespective of the type of bosonic multimode system or specifics of the experimental
approach~\cite{Ole_23_Quantumness}. At least for gaussian states
quantumness~\cite{Ole_23_Quantumness} quantifies mode entanglement better than Duan's inseparability
criterion for entanglement quantification~\cite{Duan_Cirac_Zoller__PRL00_inseparability}.

\section*{Acknowledgements}
This work is partially supported by the National Science and Technology Council of Taiwan (Nos
112-2123-M-007-001, 112-2119-M-008-007, 114-2112-M-007-044-MY3), Office of Naval Research Global,
and the collaborative research program of the Institute for Cosmic Ray Research (ICRR) at the
University of Tokyo.

\bibliography{Ole_Bibliography}

\clearpage

\setcounter{section}{0}
\renewcommand{\thesection}{SI~\arabic{section}}
\renewcommand{\thefigure}{SI.~\arabic{figure}}
\setcounter{figure}{0}
\setcounter{equation}{0}
\renewcommand{\theequation}{SI.~\arabic{equation}} 
\renewcommand{\thetable}{SI.~\arabic{table}}
\renewcommand*{\theHtable}{\thetable}
\renewcommand*{\theHfigure}{\thefigure}
\renewcommand*{\theHsection}{\thesection}
\renewcommand*{\theHequation}{\theequation}

\onecolumngrid

\begin{center}
  { \large \bf -- Supplementary Information -- \\ \vspace{0.25cm} Monitoring Beam Splitter Entanglement using Quantumness} 
\\ \vspace{0.25cm}

{{Hua-Li Chen}, {Hsien-Yi Hsieh\orcidlink{0000-0001-5227-8248}}, {Chien-Ming Wu}, {Ole
  Steuernagel\orcidlink{0000-0001-6089-7022}}, and Ray-Kuang Lee\orcidlink{0000-0002-7171-7274}}

 \end{center}

\onecolumngrid  

\setcounter{figure}{0}

\begin{figure}[h] \centering
 \includegraphics[width=12cm]{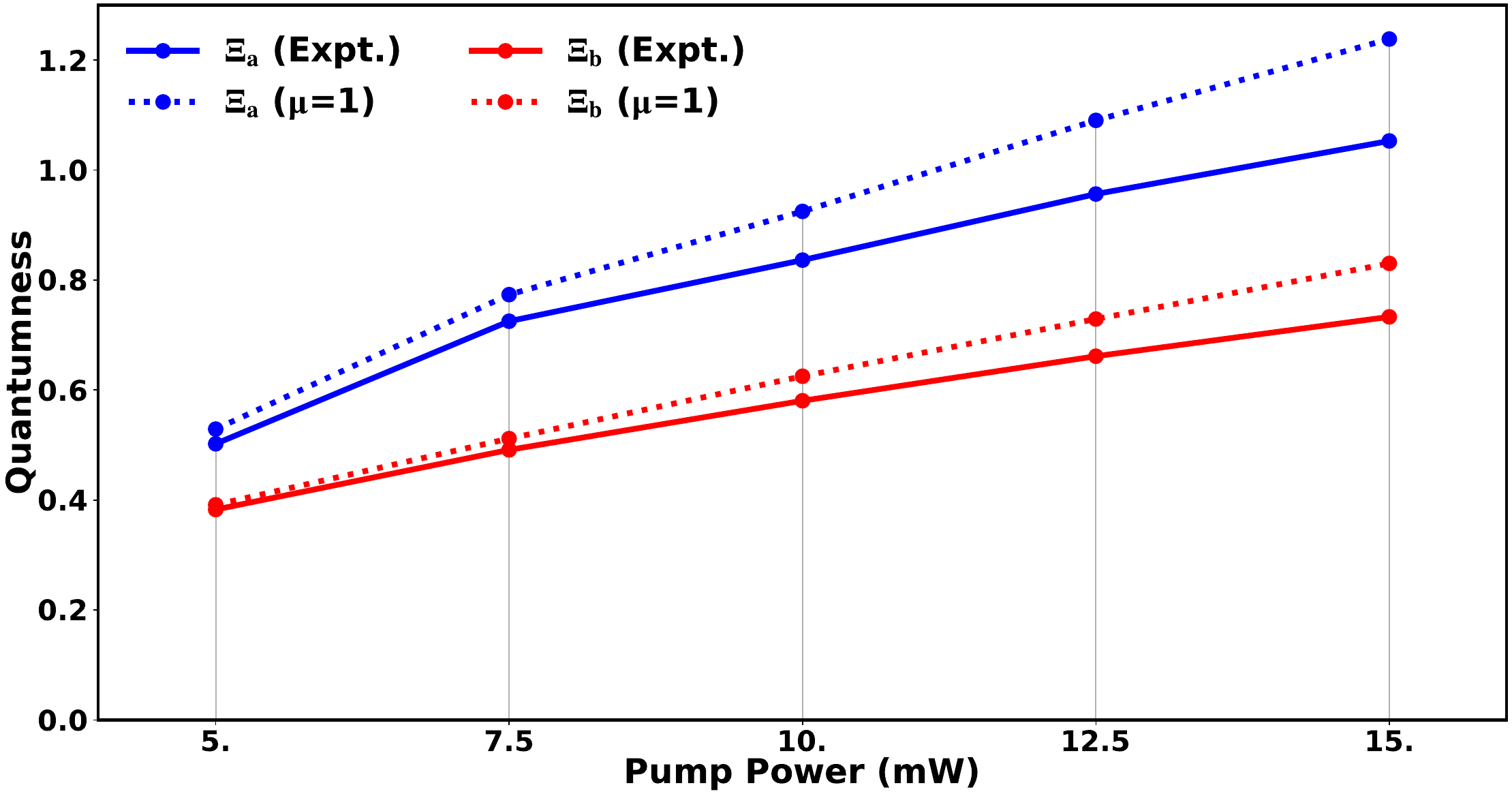}
 \caption{Quantumness for input modes: $\Xi_a$ and $\Xi_b$ of experimentally reconstructed impure
   input modes (solid curves) compared to their quantumness if all input states $W_a$ and $W_b$ were
   pure, i.e., assuming that in Eq.~(\ref{eq:W0_productState_V_mu}) $\mu_a = \mu_b \equiv \mu = 1$
   (dashed curves).}
\label{Fig:input}
\end{figure}

\begin{figure}[h]
 \includegraphics[width=8.9cm]{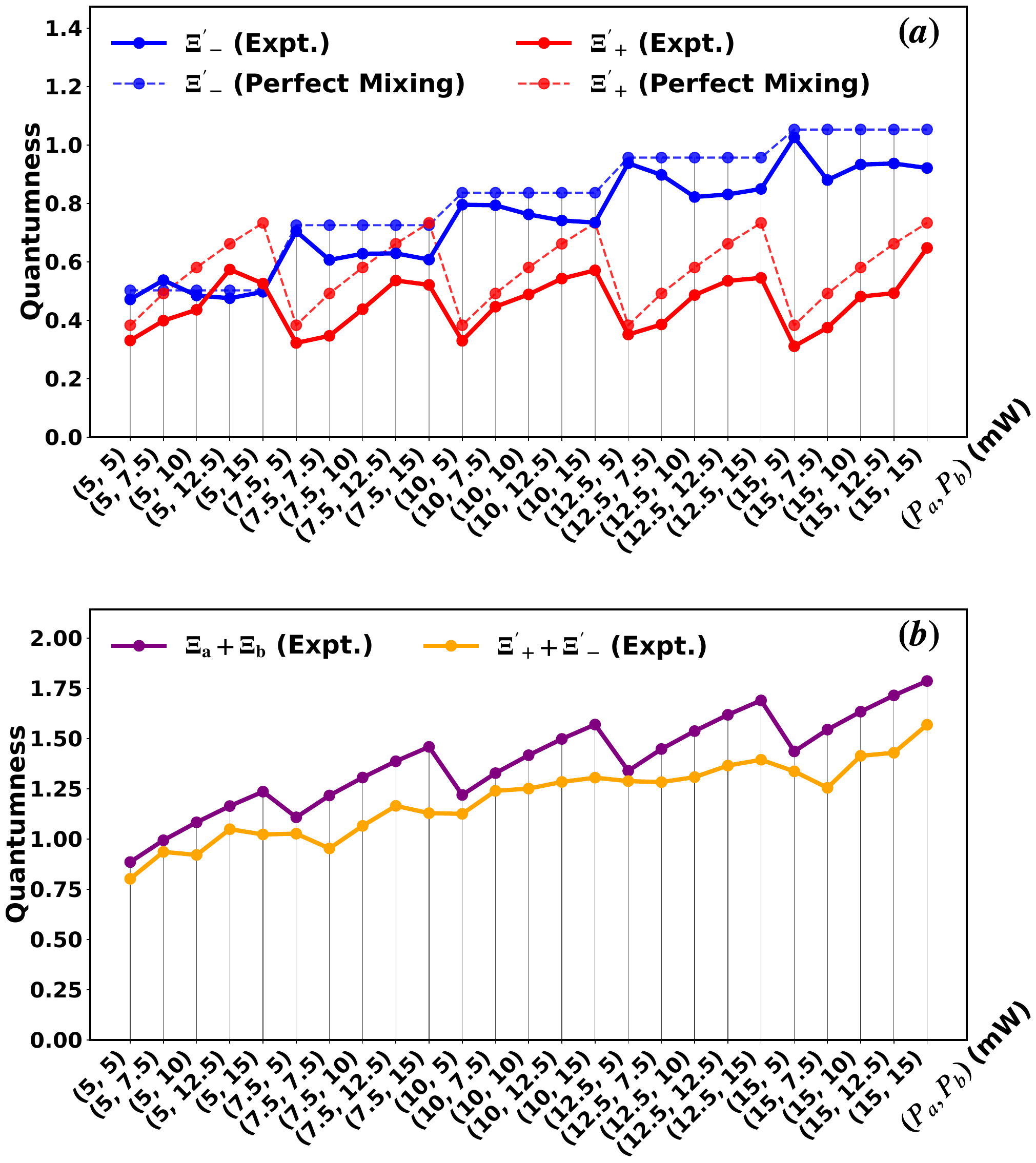} 
 \caption{Quantumness for (a) output mode: $\Xi^\prime_{+}$, $\Xi^\prime_{-}$ measured from
   experiments (solid curves), compared to the quantumness values from the Perfect Mixing (dashed
   curves), as thin lines, through the coordinate transformation with the input modes in
   Fig.~\ref{Fig:input}. The comparison between sum of quantumness of input modes $\Xi_a + \Xi_b$
   and output modes (bottom) $\Xi^\prime_{+} + \Xi^\prime_{-}$, both are experimental values, are
   shown in (b) for different pump powers of OPOs `$a$' and `$b$', in terms of ($P_a$, $P_b$).}
 \label{fig:_linePlots_before_after_BS}
\end{figure}

\newpage
\textit{Remarks on Experimental Techniques.}---As illustrated in Fig.~1, our experimental
setup consists of three bow-tie cavities: one second-harmonic generator (SHG) and two optical
parametric oscillators (OPOs), all of which operate based on the $\chi^{(2)}$-nonlinear effect of
the crystals used. The SHG cavity employs a periodically poled lithium niobate (PPLN) crystal to provide
frequency doubling, converting the laser's 1064 nm into a 532 nm beam that serves as the pump source
for the OPOs. The two OPOs, utilizing periodically poled KTiOPO4 (PPKTP) crystals, operate below the
oscillation threshold to generate squeezed vacuum states at 1064 nm. By varying the injected pump
power, squeezed vacuum states with different squeezing levels are obtained. The squeezed vacuum
states generated by the two independent OPOs are superimposed on a 50:50 beam splitter (BS) with a
relative phase of 90 degrees to prepare a two-mode squeezed vacuum (TMSV) state, which is
an entangled state. The two spatial modes of the entangled state are individually mixed
with two local oscillator beams (LOs) of equal power at two 50:50 beam splitters. The mixed optical
fields are then directed into a pair of nearly identical balanced homodyne detectors (BHDs) which are fed
into addition and subtraction circuits for joint measurements. By locking the BHDs measurement phases
at different positions, such as $x'_a$, $x'_b$, $p'_a$, and $p'_b$, the final data are acquired
using a spectrum analyzer (SA).The power of each LO is set to 12 mW. At this LO power, the two BHDs
exhibit closely matched shot-noise-to-electronic-noise clearances of 29.8 dB and 29.4 dB,
respectively. All experimental data are recorded by the SA at an analysis frequency of 2.5 MHz over
a measurement time of 10 s. The resolution bandwidth (RBW) and video bandwidth (VBW) of the SA are
set to 10 kHz and 100 Hz, respectively, with the results expressed in units of decibels (dB).

\vspace{2cm}
\begin{table}[htbp]
\centering

\resizebox{\textwidth}{!}{%
  \renewcommand{\arraystretch}{2.2}  %
  \setlength{\tabcolsep}{10.0pt}     %
  
  \begin{tabular}{|c|c|c|c|c|}
  \hline
  \multirow{2}{*}{\textbf{Pump power (mW)}} & \multicolumn{2}{c|}{\textbf{OPO `$\boldsymbol{a}$'}} & \multicolumn{2}{c|}{\textbf{OPO `$\boldsymbol{b}$'}} \\ \cline{2-5}
   & $\mathbf{Var}(\boldsymbol{x_a})$ & $\mathbf{Var}(\boldsymbol{p_a})$ & $\mathbf{Var}(\boldsymbol{x_b})$ & $\mathbf{Var}(\boldsymbol{p_b})$ \\ \hline
  \textbf{5.0}  & 2.2035& 0.5353& 0.6069& 1.8202\\ \hline
  \textbf{7.5}  & 2.7138& 0.4435& 0.5433 & 2.1116\\ \hline
  \textbf{10.0} & 3.2657& 0.4008& 0.4949& 2.501\\ \hline
  \textbf{12.5} & 3.8714& 0.3623& 0.4577& 2.8463\\ \hline
  \textbf{15.0} & 4.5173& 0.3332& 0.4265& 3.2397\\ \hline
  \end{tabular}%
}
\caption{Experimental variances of the quadratures for the two OPOs (`${a}$'
  and `${b}$') measured at different pump powers, before the beam splitter.
  }
    \label{tab:opo_data}
\end{table}

\begin{table}[p] 
\centering

\renewcommand{\arraystretch}{1.5}  
\setlength{\tabcolsep}{6.0pt}      

\resizebox{0.97\textwidth}{!}{%
\begin{tabular}{|c|c|c|c|c|c|}
\hline
\begin{tabular}[c]{@{}c@{}}\textbf{$\boldsymbol{P}_{\boldsymbol{a}}$ (mW)}\\\textbf{(OPO `$\boldsymbol{a}$')}\end{tabular}& 
\begin{tabular}[c]{@{}c@{}}\textbf{$\boldsymbol{P}_{\boldsymbol{b}}$ (mW)}\\\textbf{(OPO `$\boldsymbol{b}$')}\end{tabular}
& 
$\mathbf{Var}(\boldsymbol{x^\prime_{a} + x^\prime_{b}})$ & 
$\mathbf{Var}(\boldsymbol{p^\prime_{a} - p^\prime_{b}})$ & 
$D_c$ & 
$2 - D_c$ \\
\hline

\multirow{5}{*}{\textbf{5.0}} & \textbf{5.0} & 0.6396 & 0.5515 & 1.1911 & 0.8089 \\ \cline{2-6}
  & \textbf{7.5} & 0.5921 & 0.5231 & 1.1152 & 0.8848 \\ \cline{2-6}
  & \textbf{10.0} & 0.5591 & 0.5448 & 1.1039 & 0.8961 \\ \cline{2-6}
  & \textbf{12.5} & 0.4881 & 0.5495 & 1.0376 & 0.9624 \\ \cline{2-6}
  & \textbf{15.0} & 0.4929 & 0.5381 & 1.0310 & 0.9690 \\ \hline

\multirow{5}{*}{\textbf{7.5}} & \textbf{5.0} & 0.6456 & 0.4497 & 1.0953 & 0.9047 \\ \cline{2-6}
  & \textbf{7.5} & 0.6225 & 0.4830 & 1.1055 & 0.8945 \\ \cline{2-6}
  & \textbf{10.0} & 0.5555 & 0.4746 & 1.0302 & 0.9698 \\ \cline{2-6}
  & \textbf{12.5} & 0.5035 & 0.4740 & 0.9774 & 1.0226 \\ \cline{2-6}
  & \textbf{15.0} & 0.4948 & 0.4802 & 0.9750 & 1.0250 \\ \hline

\multirow{5}{*}{\textbf{10.0}} & \textbf{5.0} & 0.6406 & 0.4094 & 1.0500 & 0.9500 \\ \cline{2-6}
  & \textbf{7.5} & 0.5626 & 0.4117 & 0.9743 & 1.0257 \\ \cline{2-6}
  & \textbf{10.0} & 0.5315 & 0.4188 & 0.9504 & 1.0496 \\ \cline{2-6}
  & \textbf{12.5} & 0.4992 & 0.4240 & 0.9232 & 1.0768 \\ \cline{2-6}
  & \textbf{15.0} & 0.4783 & 0.4268 & 0.9051 & 1.0949 \\ \hline

\multirow{5}{*}{\textbf{12.5}} & \textbf{5.0} & 0.6266 & 0.3647 & 0.9913 & 1.0087 \\ \cline{2-6}
  & \textbf{7.5} & 0.5981 & 0.3764 & 0.9745 & 1.0255 \\ \cline{2-6}
  & \textbf{10.0} & 0.5325 & 0.3918 & 0.9243 & 1.0757 \\ \cline{2-6}
  & \textbf{12.5} & 0.5008 & 0.3870 & 0.8878 & 1.1122 \\ \cline{2-6}
  & \textbf{15.0} & 0.4870 & 0.3848 & 0.8718 & 1.1282 \\ \hline

\multirow{5}{*}{\textbf{15.0}} & \textbf{5.0} & 0.6564 & 0.3386 & 0.9950 & 1.0050 \\ \cline{2-6}
  & \textbf{7.5} & 0.5993 & 0.3673 & 0.9667 & 1.0333 \\ \cline{2-6}
  & \textbf{10.0} & 0.5345 & 0.3539 & 0.8884 & 1.1116 \\ \cline{2-6}
  & \textbf{12.5} & 0.5185 & 0.3528 & 0.8713 & 1.1287 \\ \cline{2-6}
  & \textbf{15.0} & 0.4519 & 0.3582 & 0.8100 & 1.1900 \\ \hline
\end{tabular}%
}

\caption{Experimental values of Duan's criterion from the joint quadratures measurement after the beam splitter, i.e., $D_c = \text{Var}(\hat{x}^\prime_{\boldsymbol{a}} + \hat{x}^\prime_{\boldsymbol{b}}) + \text{Var}(\hat{p}^\prime_{\boldsymbol{a}} - \hat{p}^\prime_{\boldsymbol{b}})$ in Eq.~\eqref{eq:_Duan_C}, with two independent squeezers from OPO `${a}$' and OPO `${b}$'
at different pump powers (in mW), respectively. Here, the inseparability is satisfied when $D_c < 2$.}
    \label{tab:duan_values}
\end{table}
\clearpage

\begin{table}[p] 
\centering

\renewcommand{\arraystretch}{2.2} 
\setlength{\tabcolsep}{1.5pt} 
\resizebox{0.97\textwidth}{!}{%
\begin{tabular}{|c|c|c|c|c|c|c|c|c|c|}
\hline
\begin{tabular}[c]{@{}c@{}}\textbf{$\boldsymbol{P}_{\boldsymbol{a}}$ (mW)}\\\textbf{(OPO `$\boldsymbol{a}$')}\end{tabular}& 
\begin{tabular}[c]{@{}c@{}}\textbf{$\boldsymbol{P}_{\boldsymbol{b}}$ (mW)}\\\textbf{(OPO `$\boldsymbol{b}$')}\end{tabular}
& 
$\mathbf{Var}(\boldsymbol{x^\prime_{a}})$ & $\mathbf{Var}(\boldsymbol{p^\prime_{a}})$ & 
$\mathbf{Var}(\boldsymbol{x^\prime_{b}})$ & $\mathbf{Var}(\boldsymbol{p^\prime_{b}})$ & 
$\mathbf{Var}(\boldsymbol{x^\prime_{a} + x^\prime_{b}})$ & $\mathbf{Var}(\boldsymbol{p^\prime_{a} + p^\prime_{b}})$ & 
$\mathbf{Var}(\boldsymbol{x^\prime_{a} - x^\prime_{b}})$ & $\mathbf{Var}(\boldsymbol{p^\prime_{a} - p^\prime_{b}})$ \\
\hline

\multirow{5}{*}{\textbf{5.0}} & \textbf{5.0} & 1.4142 & 1.1220 & 1.4015 & 1.1119 & 0.6396 & 1.7767 & 2.1261 & 0.5515 \\ \cline{2-10}
 & \textbf{7.5} & 1.3871 & 1.1888 & 1.4288 & 1.2485 & 0.5921 & 1.9844 & 2.1518 & 0.5231 \\ \cline{2-10}
 & \textbf{10.0} & 1.5574 & 1.2595 & 1.5434 & 1.2726 & 0.5591 & 2.3358 & 2.1484 & 0.5448 \\ \cline{2-10}
 & \textbf{12.5} & 1.7482 & 1.2120 & 1.6997 & 1.2974 & 0.4881 & 2.7718 & 2.1314 & 0.5495 \\ \cline{2-10}
 & \textbf{15.0} & 1.7824 & 1.2841 & 1.8719 & 1.3228 & 0.4929 & 3.1332 & 2.1851 & 0.5381 \\
\hline

\multirow{5}{*}{\textbf{7.5}} & \textbf{5.0} & 1.6184 & 1.0387 & 1.6997 & 1.0700 & 0.6456 & 1.7597 & 2.7018 & 0.4497 \\ \cline{2-10}
 & \textbf{7.5} & 1.6823 & 1.1660 & 1.6352 & 1.1337 & 0.6225 & 1.9512 & 2.5932 & 0.4830 \\ \cline{2-10}
 & \textbf{10.0} & 1.6184 & 1.2841 & 1.6352 & 1.4567 & 0.5555 & 2.4014 & 2.6411 & 0.4746 \\ \cline{2-10}
 & \textbf{12.5} & 1.8172 & 1.3871 & 1.7329 & 1.4848 & 0.5035 & 2.7134 & 2.6480 & 0.4740 \\ \cline{2-10}
 & \textbf{15.0} & 1.9629 & 1.4142 & 1.9453 & 1.4848 & 0.4948 & 3.1226 & 2.6694 & 0.4802 \\
\hline

\multirow{5}{*}{\textbf{10.0}} & \textbf{5.0} & 1.9253 & 1.1220 & 2.0220 & 1.1119 & 0.6406 & 1.7681 & 3.2991 & 0.4094 \\ \cline{2-10}
 & \textbf{7.5} & 1.8884 & 1.2357 & 1.9080 & 1.2974 & 0.5626 & 2.1325 & 3.2308 & 0.4117 \\ \cline{2-10}
 & \textbf{10.0} & 1.8527 & 1.4142 & 1.8719 & 1.4015 & 0.5315 & 2.4556 & 3.2483 & 0.4188 \\ \cline{2-10}
 & \textbf{12.5} & 1.9629 & 1.4418 & 1.9453 & 1.5434 & 0.4992 & 2.7683 & 3.2518 & 0.4240 \\ \cline{2-10}
 & \textbf{15.0} & 2.1203 & 1.5276 & 1.9833 & 1.6039 & 0.4783 & 3.1000 & 3.2198 & 0.4268 \\
\hline

\multirow{5}{*}{\textbf{12.5}} & \textbf{5.0} & 2.2034 & 1.0590 & 2.3141 & 1.0495 & 0.6266 & 1.7869 & 3.9325 & 0.3647 \\ \cline{2-10}
 & \textbf{7.5} & 2.0797 & 1.1439 & 2.2263 & 1.2246 & 0.5981 & 2.0053 & 3.7499 & 0.3764 \\ \cline{2-10}
 & \textbf{10.0} & 2.1203 & 1.4142 & 2.1423 & 1.4015 & 0.5325 & 2.4550 & 3.7781 & 0.3918 \\ \cline{2-10}
 & \textbf{12.5} & 2.1617 & 1.5574 & 2.2698 & 1.5735 & 0.5008 & 2.8012 & 3.9074 & 0.3870 \\ \cline{2-10}
 & \textbf{15.0} & 2.2034 & 1.5874 & 2.2263 & 1.6352 & 0.4870 & 3.1023 & 3.8268 & 0.3848 \\
\hline

\multirow{5}{*}{\textbf{15.0}} & \textbf{5.0} & 2.5223 & 1.0000& 2.5976 & 1.0000& 0.6564 & 1.6694 & 4.4567 & 0.3386 \\ \cline{2-10}
 & \textbf{7.5} & 2.4266 & 1.2595 & 2.5485 & 1.2485 & 0.5993 & 2.1113 & 4.3575 & 0.3673 \\ \cline{2-10}
 & \textbf{10.0} & 2.4740 & 1.3871 & 2.4518 & 1.3746 & 0.5345 & 2.4578 & 4.5273 & 0.3539 \\ \cline{2-10}
 & \textbf{12.5} & 2.4740 & 1.4983 & 2.5485 & 1.5735 & 0.5185 & 2.7534 & 4.5457 & 0.3528 \\ \cline{2-10}
 & \textbf{15.0} & 2.4266 & 1.7151 & 2.5976 & 1.7664 & 0.4519 & 3.1770 & 4.4132 & 0.3582 \\
\hline
\end{tabular}%
}
\caption{Experimental variances of the individual and joint quadratures for modes $a$
  and $b$ measured after the 50:50 beam splitter (BS) under various combinations of
  pump powers ($P_{\boldsymbol{b}}$ and $P_{\boldsymbol{a}}$). The table displays the 
  quadrature variances ($\text{Var}(x^\prime_{\boldsymbol{a}})$,
  $\text{Var}(p^\prime_{\boldsymbol{a}})$, $\text{Var}(x^\prime_{\boldsymbol{b}})$,
  $\text{Var}(p^\prime_{\boldsymbol{b}})$) along with the joint variances
  $\text{Var}(x^\prime_{\boldsymbol{a}} \pm x^\prime_{\boldsymbol{b}})$ and
  $\text{Var}(p^\prime_{\boldsymbol{a}} \pm p^\prime_{\boldsymbol{b}}$)) used to determine
  quantum correlation properties and entanglement of the output states.}
    \label{tab:detailed_joint_data}
\end{table}
\clearpage


\end{document}